\documentclass[aps,nofootinbib]{revtex4}
\usepackage{epsfig}
\begin{document}
\title{Scaling theory of excitations in trapped fermionic condensates}
\author{G.F.~Bertsch$^{1,2}$, A.~Bulgac$^{1}$, and R.A.~Broglia$^{3,4}$ }
\affiliation{
$^{1}$Department of Physics and $^{2}$Institute of Nuclear Theory,
University of Washington,
Seattle, USA\\
and $^3$Department of Physics University of Milan, Milan, Italy\\
$^4$INFN Sezione di Milano, Milan, Italy}

\def\be{\begin{equation}}
\def\ee{\end{equation}}
\def\<{\langle}
\def\>{\rangle}
\begin{abstract}
Formulas are derived for the coupled quadrupolar and monopolar
oscillations of a fermion condensate trapped in a axially
symmetric harmonic potential. We consider two-component condensates
with a large particle-particle scattering length and in the limits of
frequencies either small (adiabatic) or large (diabatic) compared to the
pairing gap.  Using the method of coordinate
scaling, we reproduce Stringari's results for adiabatic
excitations and present new results for the diabatic limit.
Unlike the adiabatic oscillations, the diabatic ones show sensitivity
to the interaction between particles.

\end{abstract}
\maketitle

\section{Introduction}

Experimental techniques in atomic cooling and trapping have now
evolved to a point where condensates of fermionic atoms such as
$^6$Li can be produced and studied \cite{experiments}.
One of the interesting measurable properties is the spectrum of 
collective excitations.  Stringari \cite{st03} has investigated this question
for the case of the unitary limit of a two-component
system.  By ``unitary" is meant that the scattering length associated
with the two-body interaction is large compared to all other scales in
system. In ref. \cite{st03} it is assumed that the system behaves 
hydrodynamically.  This requires the dynamics to maintain an
isotropic pressure tensor when the system is deformed.  The 
condition for this adiabatic limit to apply to fermion systems
is that they be superfluid and that the collective oscillation frequency
be much less than the pairing gap.  The
other extreme, the diabatic limit, obtains if the oscillation frequency
is much greater than the pairing gap.  This in fact is the case for
the giant resonances in nuclear spectroscopy.  In that limit, 
the dynamics is governed by the mean field acting on the
particle orbitals and coupling through the ordinary density.  We believe
it is of interest to investigate this limit for the atomic condensates
as well, to understand the dependence of the frequencies on all
possible parameters of the system.  

The time-dependent mean-field theory may solved in the
small-amplitude limit by the RPA equations, but inclusion 
of all the single-particle degrees of freedoms requires a considerable
numerical effort.  This is especially the case for the theory including
the pairing degrees of freedom as well\cite{br01,br02}. Much simpler analytic 
formulas can be derived under the assumption that
the excitations are fully collective, meaning that they exhaust
the sum rule for some one-body field.  In the sum rule approach,
the formulas express the oscillation frequency in terms of
a  ratio of two integrals involving only ground-state quantities
in the integrands.  Again referring to the
experience in nuclear physics, this approach has been very  
successful, particularly with the monopole and quadrupole modes
of oscillation.  

Considering atomic condensates in an axially symmetric harmonic
trap, it is natural to consider compressional and quadrupolar
modes generated by the fields   $Q = x^2,y^2,$ and $z^2$.
These fields induce scaling transformations on the orbital wave
functions, e.g., for the field $x^2$, the transformed 
wave function is \cite{be97}
\be
\phi(x,y,z) \rightarrow e^{\beta_x [H,x^2]/2}\phi = e^{-\beta_x/2} 
\phi(e^{-\beta_x} x, y, z)
\ee
where $\beta_x$ is a scaling coordinate.
The collective frequencies may be calculated by constructing
the energy function in the scaling coordinates and solving the
classical equation of motion.  This route is rather transparent
and we follow it in the derivation below.  Consider an energy 
function of the form  
\be
E = T + V_{ext} + U
\ee
where $T$ is the kinetic energy operator, $U$ is the interaction
energy, and $V_{ext}$ is the trap field.  We write the trap field
as
\be
V_{ext} = {1\over 2}\Big( x^2 + y^2 + (\lambda z)^2\Big), 
\ee
setting the mass of the particles and 
the transverse harmonic oscillator frequency to one.
In the Thomas-Fermi approximation of a free fermion gas
the ground state density is given by
\be
\label{rho}
\rho_0(\vec r) = c \Big(1- (x/R)^2 -(y/R)^2 -(\lambda z / R)^2\Big)^{3/2},
\ee
where $c$ is a constant and $R$ is the transverse radius of the
condensate. The kinetic energy is given by 
\be
\<T\>_{TF} = \int d^3r\, \rho_0^{5/3} {3\over 10}\left({3 \pi^2  }\right)^{2/3}
\ee
in a two-component Fermi gas.
The density also has the form of eq.~(\ref{rho}) in an interacting two-component fermion
system in the unitary limit.
The scaled density is given by
\def\bx{\beta_x}
\def\by{\beta_y}
\def\bz{\beta_z}
\be
\rho_{\beta}(\vec r) = c e^{-\bx-\by-\bz} \Big(1- (e^{-\bx}x/ R)^2 -
(e^{-\by}y/R)^2-( \lambda e^{-\bz}z / R)^2\Big)^{3/2}.
\ee
We next evaluate the various terms in the energy as a function of
the scaling variables.  The expectation of the external potential is
given by
\be
\langle V_{ext} \rangle = a ( e^{2\bx}+e^{2\by}+e^{2\bz})
\ee
where the prefactor $a$ is the integral $a = \int d^3 r \rho_0 x^2/2$.
The kinetic energy may vary in two different ways, depending on whether 
or not the
Fermi surface remains spherical under the dynamic evolution.  
In the adiabatic limit we have  
\be
\label{adiabatic}
\<T\>_{ad} =  b e^{-2(\bx+\by+\bz)/3}
\ee
where $b = \<T\>_{TF}$. In
the diabatic limit the contributions in the three Cartesian directions are
independent and the kinetic energy is
\be
\<T\>_{dia} = {b  \over 3} (e^{-2\bx}+e^{-2\by}+e^{-2\bz}).
\ee  
We next turn to the potential energy.  It behaves very simply in the
unitary limit, being proportional to the adiabatic kinetic energy with a 
universal constant of proportionality. We shall further assume
that the potential energy
behave as a scalar, creating an isotropic pressure.  
It thus depends only on $\<T\>_{ad}$.  We parameterize it as
\be
\label{unitary}
\<U\> = (\xi -1) \<T\>_{ad}
\ee

We may now calculate the vibrations in a harmonic approximation
making a second-order Taylor expansion of the energy function of the 
collective variables.   The restoring force constants are given by the
hessian matrix {\bf K}, with
\be
\label{hessian}
K_{ij}  = {\partial^2 \<E\>\over \partial \beta_i \partial \beta_j}.
\ee
We will also need the inertia tensor {\bf I} associated the collective
variables.  Defining the velocity field $\vec v = \vec\beta \nabla Q/2$, the element
$I_{ij}$ is the coefficient of $\dot \beta_i \dot \beta_j$ in the expression
$\int d^3 r \rho_0 v\cdot v$.  In our case the integral  is
\be
I_{ij} = \int d^3r \rho_0  r_i r_j.
\ee
Note that the integral is essentially identical to that for $\<V_{ext}\>$. 
Thus
\be
{\bf I} = 2 a\left( \begin{array}{ccc} 
1 & 0 & 0  \\
0 & 1 & 0  \\
0 & 0 & \lambda^{-2} \\
\end{array} \right). 
\ee
\be
{\bf K } u   = \omega^2 {\bf I}  u
\ee

Before applying the formulas for the diabatic frequencies, we 
confirm previous results on the adiabatic frequencies\cite{st03,am99}.  Here we
use eq. (\ref{adiabatic}) for the kinetic energy and eq. (\ref{unitary})
for the interaction energy, i.e. setting $\<E\> = \xi \< T\>_{ad}$ in 
eq.~(11).
We also use the fact that $\vec \beta = 0 $ 
in equilibrium to eliminate the
coefficient $b$ in the expression for the kinetic energy.  The
equilibrium condition
\be
\label{equilibrium}
{\partial \<E\> \over  \partial \beta_i }|_0=0
\ee
is satisfied by setting $b = 3 a/\xi$.  The squared frequencies $\omega^2$
may then be obtained from the eigenvalues of the matrix
\be
\label{Ka}
{\bf I}^{-1/2} {\bf K}_{ad} {\bf I}^ {-1/2}= {2 \over 3 } \left( \begin{array}{ccc} 
4 & 1 & \lambda \\
1 &4 & \lambda  \\
\lambda & \lambda & 4\lambda^2 \\
\end{array} \right).
\ee
Note that all dependence on the interaction strength $\xi$ has
disappeared.  The eigenvalues of eq.~(\ref{Ka}) 
are
\be
\label{omegaa}
\omega_{1,2}^2 =(5 + 4 \lambda^2 \pm \sqrt{25-32\lambda^2+16\lambda^4})/3
\ee
$$
\omega_3^2 = 2.
$$
This agrees  ref. \cite{st03} as well as the frequencies calculated
for noninteracting fermions in ref. \cite{am99}.  To get more insight into the physics, let us specialize
to the case of a spherical trap ($\lambda = 1$).
The frequencies become  $\omega_{1}^2= 4$, corresponding to the
monopole vibration and $\omega_{2,3}^2=2$, corresponding to two of the
five degenerate quadrupolar
vibrations\footnote{It is amusing to note that the diabatic harmonic 
formula for the
the giant nuclear quadrupole resonance is also reduced from the nominal
value of $\omega^2 = 4 $  by a factor of two\cite{su73,be94}.  In both cases,
the flow is incompressible so that the co-moving density remains the
same during the oscillation.  In the nuclear case with a short-range
interaction, this implies a vanishing interaction contribution to the
restoring force constant, leaving a kinetic contribution.  The
situation is converse for the trapped fermion case:  here the
kinetic contribution vanishes leaving only the interaction with
the external field.}.  

Next we carry out the same calculation using the diabatic energy.
The self-consistency condition eq. (\ref{equilibrium}) together with the
strong-coupling relation eq. (\ref{unitary}) determines the internal energy
function to be
\be
\<T\>_{dia}  + \<U\> = {a \over  \xi} (e^{-2\bx}+e^{-2\by}+e^{-2\bz})
+{ 3 a(\xi -1)  \over \xi} e^{-2(\bx+\by+\bz)/3}.
\ee
Evaluation of the derivatives for the force matrix then yields
\be
{\bf K}_{dia} = {4 a \over 3 \xi }\left( \begin{array}{ccc} 
2 & -1 & -1 \\
-1 & 2 & -1  \\
-1& -1 & 2  \\
\end{array} \right)+{4 a\over 3 }\left( \begin{array}{ccc} 
4 & 1 & 1  \\
1 & 4 & 1   \\
1 & 1 & 4  \\
\end{array} \right).
\ee
The eigenvalues of $ {\bf I}^{-1/2} {\bf K}_{dia} {\bf I}^{-1/2} $ are
\be
\label{w-diabatic}
\omega_{1,2}^2 = (f \pm
\sqrt{f^2 -72 \xi \lambda^2 (1 + \xi)})/3 \xi
\ee
$$
\omega_3^2 = 2 + {2\over \xi}
$$
where $f= 1 + 2 \lambda^2 + 5 \xi + 4 \lambda^2 \xi$.  
For the convenience of the reader, a script is given in \cite{gb04}
of the mathematical operations leading to eqs. (\ref{omegaa},
\ref{w-diabatic}).

The frequencies in eq. (\ref{w-diabatic}) depend on the interaction 
constant $\xi$, in contrast to the adiabatic frequencies.  Their measurement
would thus provide information about the adiabaticity and thus about
the relative size of the pairing gap and the vibrational frequency.  
As a final remark, let us specialize again to the spherical trap.  The
frequencies become  $\omega_1^2 = 4$ and $\omega_{2,3}^2 = 2 + 2/\xi$.
The monopole frequency $\omega_1$ is the same as in the adiabatic formula, which
is not surprising because the Fermi surface remains spherical under
the compression by the monopole scaling field.  The quadrupole modes
depend on the interaction and lie above the monopole at the physical
value of $\xi$, calculated to be $\xi\approx 0.44 $ 
with pairing and $\xi\approx 0.55 $ without pairing \cite{carlson}.
This is completely at variance which the
behavior of the quadrupole mode in other systems such as atomic
nuclei, where the incompressibility of the medium produces a
higher monopole frequency.

\section*{Acknowledgments}

This work was supported  in part by the Department of Energy under grants
DE-FG06-90ER40561 and DE-FG03-97ER41014.  The work was stimulated by 
discussions with R. Grimm, C. Chin, and at the workshop ``Bose-Einstein condensation 
of Fermions" at the University of Milan.

\def\etal{et al.,}

\end{document}